\documentclass[journal=mamobx,manuscript=article,layout=twocolumn]{achemso}
\usepackage{multirow}
\usepackage{lineno}
\usepackage[font=footnotesize]{subfig}
\usepackage[format=plain,
            labelfont={it, bf},
            textfont=it]{caption}
\usepackage{amsmath,amssymb}
\usepackage{graphicx}
\usepackage{xcolor}

\author{Anna Dmochowska}
\affiliation[PIMM]{Laboratoire PIMM, CNRS, Arts et Métiers Institute of Technology, Cnam, HESAM Université, 75013 Paris, France}
\author{Jorge Peixinho}
\affiliation[PIMM]{Laboratoire PIMM, CNRS, Arts et Métiers Institute of Technology, Cnam, HESAM Université, 75013 Paris, France}
\email{jorge.peixinho@cnrs.fr}
\author{Cyrille Sollogoub}
\affiliation[PIMM]{Laboratoire PIMM, CNRS, Arts et Métiers Institute of Technology, Cnam, HESAM Université, 75013 Paris, France}
\author{Guillaume Miquelard-Garnier}
\affiliation[PIMM]{Laboratoire PIMM, CNRS, Arts et Métiers Institute of Technology, Cnam, HESAM Université, 75013 Paris, France}
\email{guillaume.miquelardgarnier@lecnam.net}

\title{Dewetting dynamics of sheared thin polymer films: an experimental study
\footnote{This document is the unedited Author's version of a Submitted Work that was subsequently accepted for publication in ACS Macro Letters, copyright © 2022 American Chemical Society after peer review. To access the final edited and published work see https://doi.org/10.1021/acsmacrolett.2c00070}}
\keywords{thin films, polymer films, dewetting, shear}

\begin{document}

\begin{tocentry}
\includegraphics[height=4.45cm]{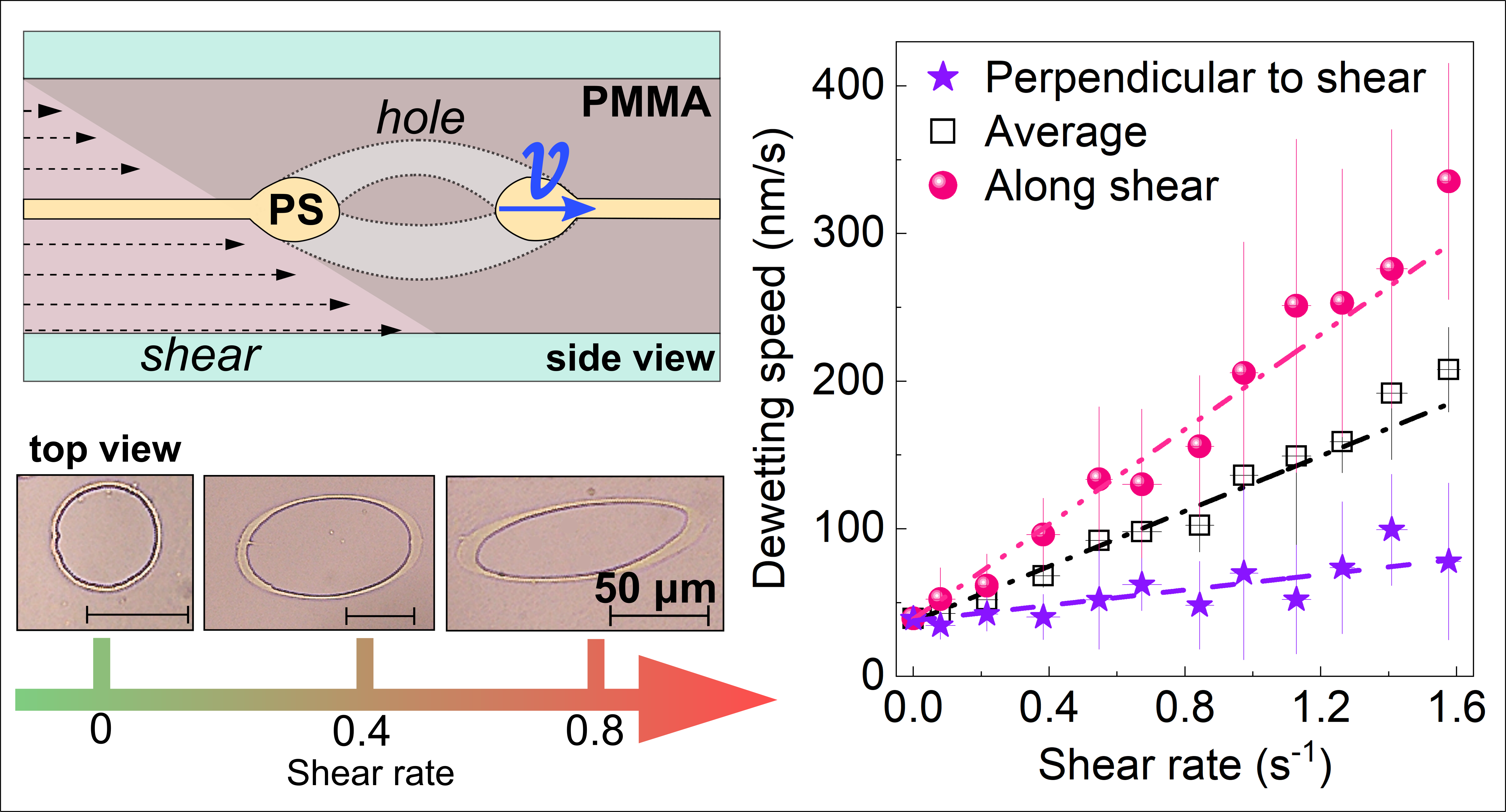}
\end{tocentry}

\begin{abstract}
An experimental investigation is reported on the effect of shear on the bursting of molten ultra-thin polymer films embedded in an immiscible matrix. 
By using an optical microscope coupled with a shearing hotstage, the dewetting dynamics, i.e. the growth of dewetting holes is monitored over time at various shear rates.
It is observed that their circularity is modified by shear and that for all temperatures and thicknesses studied, the growth speed of the formed holes rapidly increases with increasing shear rate.
A model balancing capillary forces and viscous dissipation while taking into account shear-thinning is then proposed and captures the main features of the experimental data, such as the ellipsoid shape of the holes and the faster dynamics in the direction parallel to the shear.
This research will help to understand the instabilities occurring during processing of layered polymeric structures, such as multilayer coextrusion. 
\end{abstract}

\section{}
Rupture of thin liquid films, whether free or supported on a solid or liquid substrate, has long been studied \cite{Vrij1966, Sheludko1967, Ruckenstein1974}. It can be attributed to the amplification of thermal fluctuations at the interface by long-range interactions or to the nucleation and growth of dry patches, leading to the formation of holes in the film \cite{Oron1997, Craster2009, MorenoBoza2020, Kheshgi1991}. In the supported case, while retracting, the film generally pushes material into a rim around the hole which grows due to surface tension, until the rims coalesce to form droplets. The speed at which the holes grow (i.e. dewetting speed) results from an equilibrium between surface tension that draws the rim and viscous dissipation, and depends on the nature of the substrate and the viscosity of the film.\\ 
Similar mechanisms occur in ultra-thin polymer films (thicknesses well-below 1 $\mu$m) when heated above their glass transition (or melting) temperature. They also have been thoroughly examined both from a theoretical and experimental point of view \cite{Krausch1997,BrochardWyart1993,Qu1997,Reiter1992,Vilmin2006}. The effects of viscoelasticity \cite{Mulama2019}, slip \cite{Peschka2019} or residual stresses \cite{Reiter2005} have been considered, as such films find many industrial applications in conformal, protective, optical or functional coatings\cite{Krogman2005,Katsumata2018,Runser2019,Lynge2011} and electrical films \cite{Shi1996}.\\
A case less studied is the stability of multilayer structures, i.e. stratified thin liquid (or polymeric) films of different nature \cite{Bertin2021}. They can be found in microfluidic applications \cite{Lenz2007}, packaging applications\cite{Langhe2016} or used for the fabrication of functionalized fibers \cite{Nguyen2017}, and obtained through lithographic printing process \cite{Lenz2007}, thermal drawing \cite{Nguyen2017}, layer-by-layer deposition \cite{Park2018}, or multilayer coextrusion \cite{Beuguel2019}.\\ 
Theoretical and numerical studies have evaluated the effect of thicknesses and viscosities in the modes of rupture in confined multilayers \cite{Lenz2007, Xu2020}. In recent works, aiming to elucidate the physical mechanisms responsible for layer break-ups during multilayer coextrusion \cite{Bironeau2017,Feng2018}, we developed an experimental set-up to analyze the dewetting speed in a confined trilayer \cite{Zhu2016,Chebil2018}. 
This consisted of an ultra-thin ($e\sim100$ nm) polystyrene (PS) layer in between two micrometric poly(methyl methacrylate) (PMMA) layers. Basically, the holes' growth speed, $v$, can be described as the bursting of a Newtonian fluid film in a viscous environment \cite{Reyssat2006} with dissipation occurring not in the PS film but in the surrounding PMMA. In this case $v\sim\sigma/\eta_0=v_0$, where $\sigma$ is the PS/PMMA interfacial tension, and $\eta_0$ the Newtonian PMMA viscosity. 
\par

Another barely studied question of interest is the role of shear on the dewetting. Shear is indeed applied in most of the processes used to produce these stratified structures, and understanding its effect on the stability and dewetting dynamics is then of fundamental importance. Numerical works in a bilayer flow \cite{Kalpathy2010}, for a thin liquid film \cite{Davis2010} or an ultra-thin polymer film \cite{Kadri2021} have shown that shear can delay or, above a critical shear rate, even suppress the rupture. From an experimental point of view, to the best of our knowledge, only two qualitative studies were conducted on sheared trilayer systems, but were mostly focused on the observation of the appearance of the holes \cite{Sundararaj1995,Lyngaae1996}. \par

Since it appears no work aims at elucidating the role of shear in the evolution of the dewetting dynamics, we propose here such quantitative study in a PMMA/PS/PMMA trilayer.

\begin{figure}[hbt!]
  \centering
  \includegraphics[width=8.4cm]{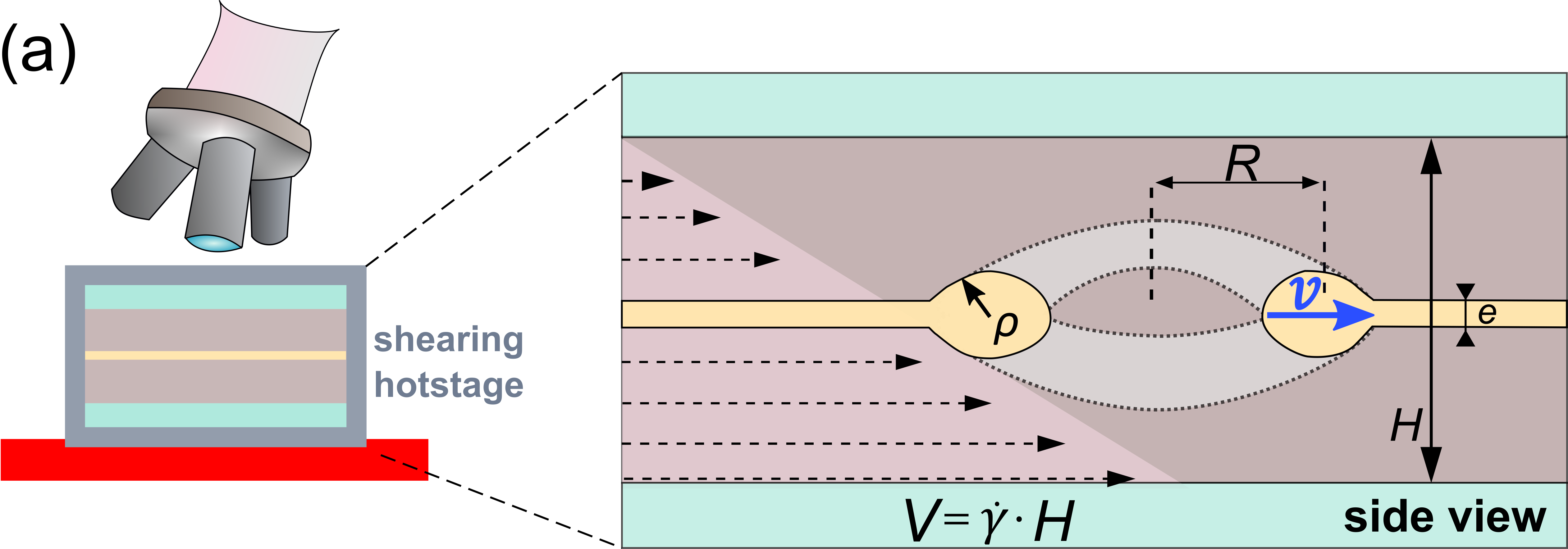}
  \hfil
  \includegraphics[width=8.4cm]{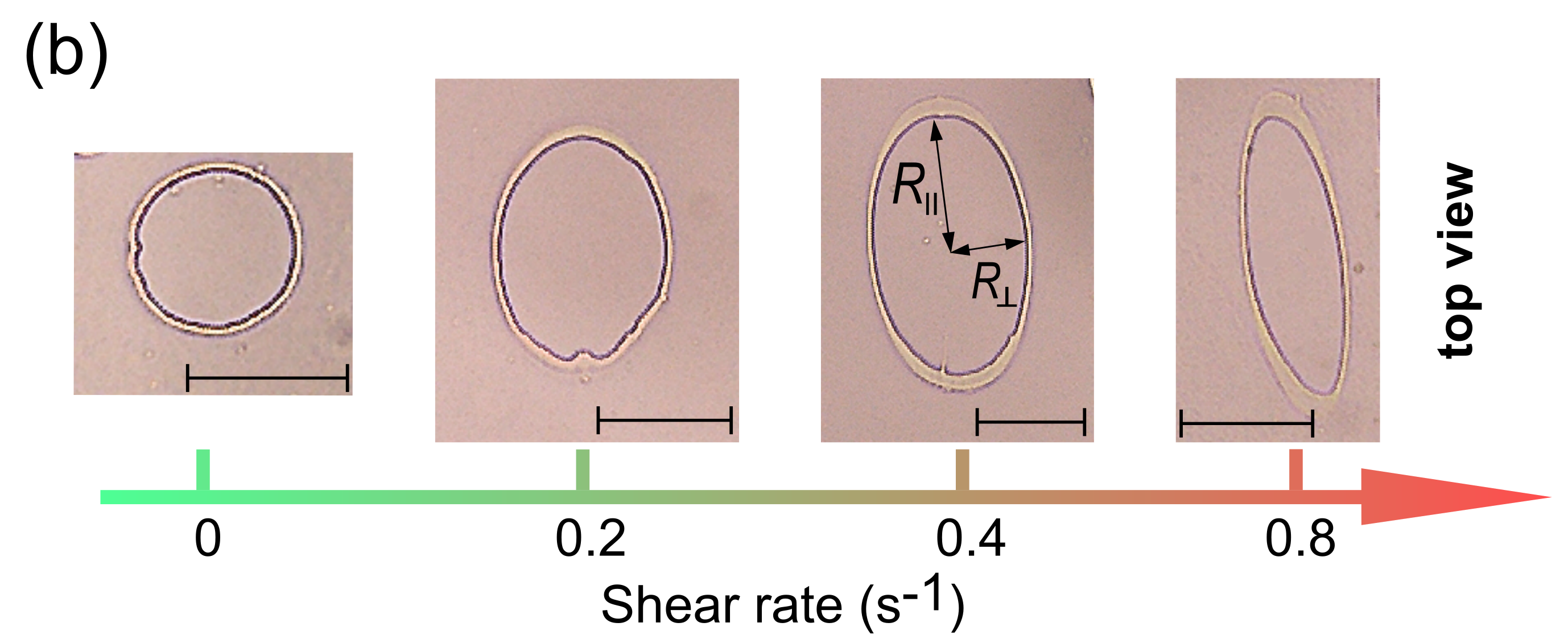}
  \hfil
  \caption{(a) Schematics of the dewetting experiment with a zoom around a hole (side view). The PS layer is embedded between two PMMA layers with same thicknesses. The top plate is fixed, the bottom plate is moving with a speed $V$ which value is set by the shear rate and the gap (or thickness of the trilayer) $H\sim 200\,\mu m$. 
  The shear rate is then defined as $\dot{\gamma}=V/H$.
  Hence, the PS layer and the holes, while growing, are moving at $V/2$ parallel to the shear.
  The dewetting of the film with thickness $e$ is observed from the top. (b) Changes in the holes morphology with increasing shear rate (top view). The scale bars are 50 $\mu$m. The holes are circular with a radius $R$ when dewetting occurs without shear, and have an ellipsoid shape when dewetting happens under shear, with a major radius $R_\parallel$ in the shear direction and a minor radius $R_\bot$ perpendicular to shear. The holes appear more elongated when higher shear is applied. The final dewetting stage is a series of long and thin holes all aligned in the shear direction.}
  \label{fig:one}
\end{figure}

First, the morphology of the holes varies significantly with shear rate, from a round shape at rest to an elongated (ellipsoid) one in the shear direction (see Figure \ref{fig:one}), with an effect more pronounced as shear increases. 
Note that the shape of the rim also seems altered, with most of its volume gathered around the vertices of the holes instead of a regular shape around the hole. A similar change in the shape of the hole was also noticed in the case of PS dewetting from a stretched elastomer substrate\cite{Schulman2018}.

We can describe the steady-state hole's growth speed $v$ without shear (i.e. when the hole is circular) based on our previous studies \cite{Zhu2016,Chebil2018}. It is defined as $R=vt$, where $R$ is the radius of the hole and $t$ the time of the experiment. We assume that after a transient regime leading to the formation of a toroidal rim, the PS initially contained in the hole's volume is gathered into it\cite{Brochard1997}.

The change of capillary energy $\mathcal{U}$ may then be written as:
\begin{equation}
\frac{d\mathcal{U}}{dt}\sim{4\sigma Rv}
\label{eq1}
\end{equation}
where $\sigma$ is the PMMA/PS interfacial tension in J/m$^2$ determined using Wu's empirical equation\cite{Wu1970} $\sigma=3.2-0.013(T-20)$ with $T$ the temperature in \textdegree C.\\
In the present study, since $H\sim1000e$, confinement effects on the dewetting dynamics can be neglected \cite{Chebil2018} and in this range of viscosity ratio $\mathcal{O}(1)$, the dissipation shall occur within the surrounding PMMA \cite{Chebil2018, Zhu2016}. Consequently, the dissipated power can be set up as:
\begin{equation}
\mathcal{D}\sim{\eta_0\rho^2R\left(\frac{v}{\rho}\right)^2 = \eta_0Rv^2}
\label{eq2}
\end{equation}
where $\eta_0$ is the PMMA Newtonian viscosity at the study temperature determined from the Carreau-Yasuda model (see Experimental Section) and $\rho$ is the radius of the rim's cross-section.

Balancing equations (\ref{eq1}) and (\ref{eq2}), a constant growth speed is obtained in the case of dewetting without external shear:
\begin{equation}
v=A_0\frac{\sigma}{\eta_0}=A_0 v_0
\label{eq3}
\end{equation}

where $A_0$ is a numerical prefactor of order unity due to geometric simplifications from the scaling approach. For the five configurations ($e,T$) tested, we obtain $A_0=1.3\pm0.3$.

\begin{figure}[t!]
  \centering
  \includegraphics[width=8.4cm]{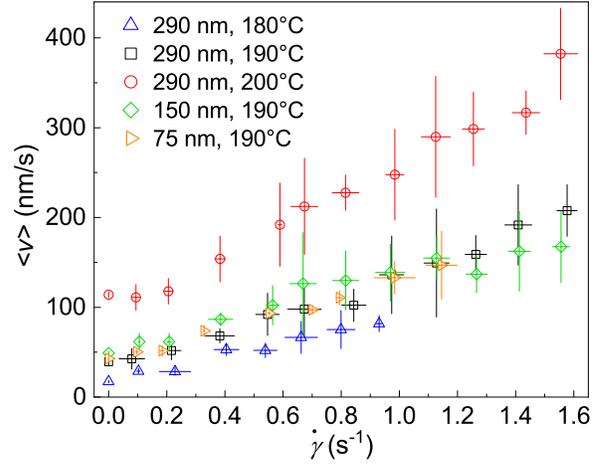}
  \hfil
  \caption{Dependence of the average growth speed on the shear rate for different $e$ and $T$. From the experimental setup, a shear rate value was obtained for each hole by measuring its moving speed. At least 5 different holes with very similar shear rates (intervals of 0.15 s$^{-1}$) were observed. The radius of the hole $\mathcal{R}$ was calculated as an average from $R_\parallel$ and $R_\bot$. The holes formed at various times of the experiment and only holes with radius of at least 5 $\mu$m  were considered, hence in the steady-state regime.  For each hole, minimum 5 images were analyzed by ImageJ software to capture the growth evolution over time. The speed for each hole was obtained from a linear regression of $\mathcal{R}$ on $t$, and \textless$v$\textgreater{} is the average from all the holes' growth speed in their given interval.}
  \label{fig:two}
\end{figure}

As a first simple comparison with the case without shear, we define an effective radius for the sheared ellipsoidal holes, such as $\mathcal{R}=\left( R_{\parallel}+R_\bot\right )/2$, where $R_{\parallel}$ is the semi-major axis and $R_\bot$ the semi-minor axis of the ellipsoid. At a given shear rate, monitoring the evolution of $\mathcal{R}$ over time leads to the average growth speed $<$$v$$>$, similar to the previous approach (see Figure \ref{fig:two}). $<$$v$$>$ increases sharply with increasing shear rate, for all tested conditions. Temperature variation has a bigger effect on the dynamics than changing the PS thickness, suggesting that viscosity still plays a major role on the dewetting. The increase of $v$ is most pronounced at 200\textdegree C, from about 100 nm/s without shear to 400 nm/s at shear rate 1.6 s$^{-1}$.\par

Considering the effect of shear on both the increasing average growth speed of the holes and the differences in their morphology, we now try to go further and propose a model describing these observations. 
In the following, a more quantitative approach will consist in evaluating separately the growth speed in the shear direction, $v_\parallel$, and perpendicular to shear, $v_\bot$, from the temporal evolution of $R_{\parallel}$ and $R_\bot$ respectively.\

\begin{figure*}[ht!]
  \centering
  \includegraphics[width=8.8cm]{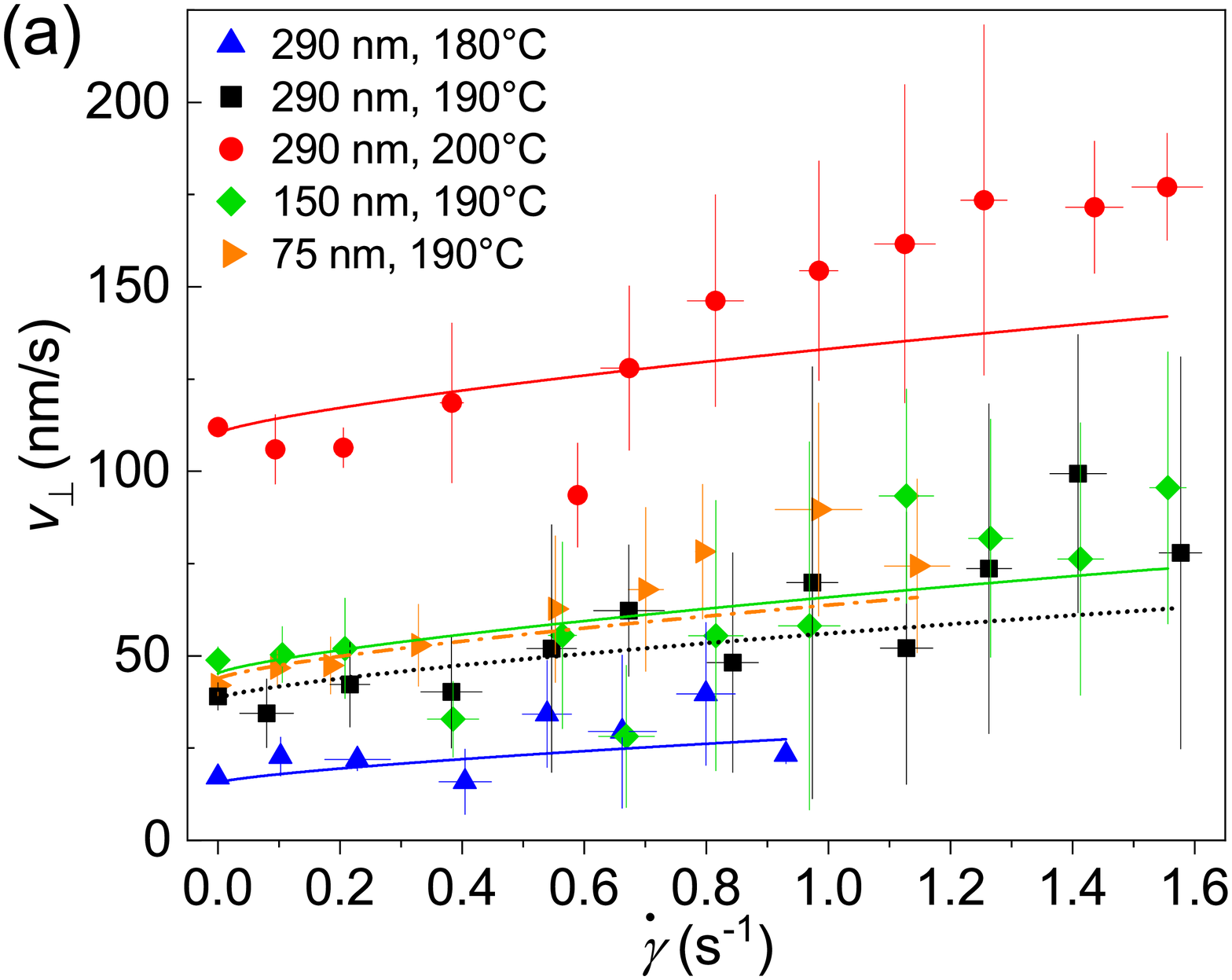}
  \hfill
  \includegraphics[width=8.8cm]{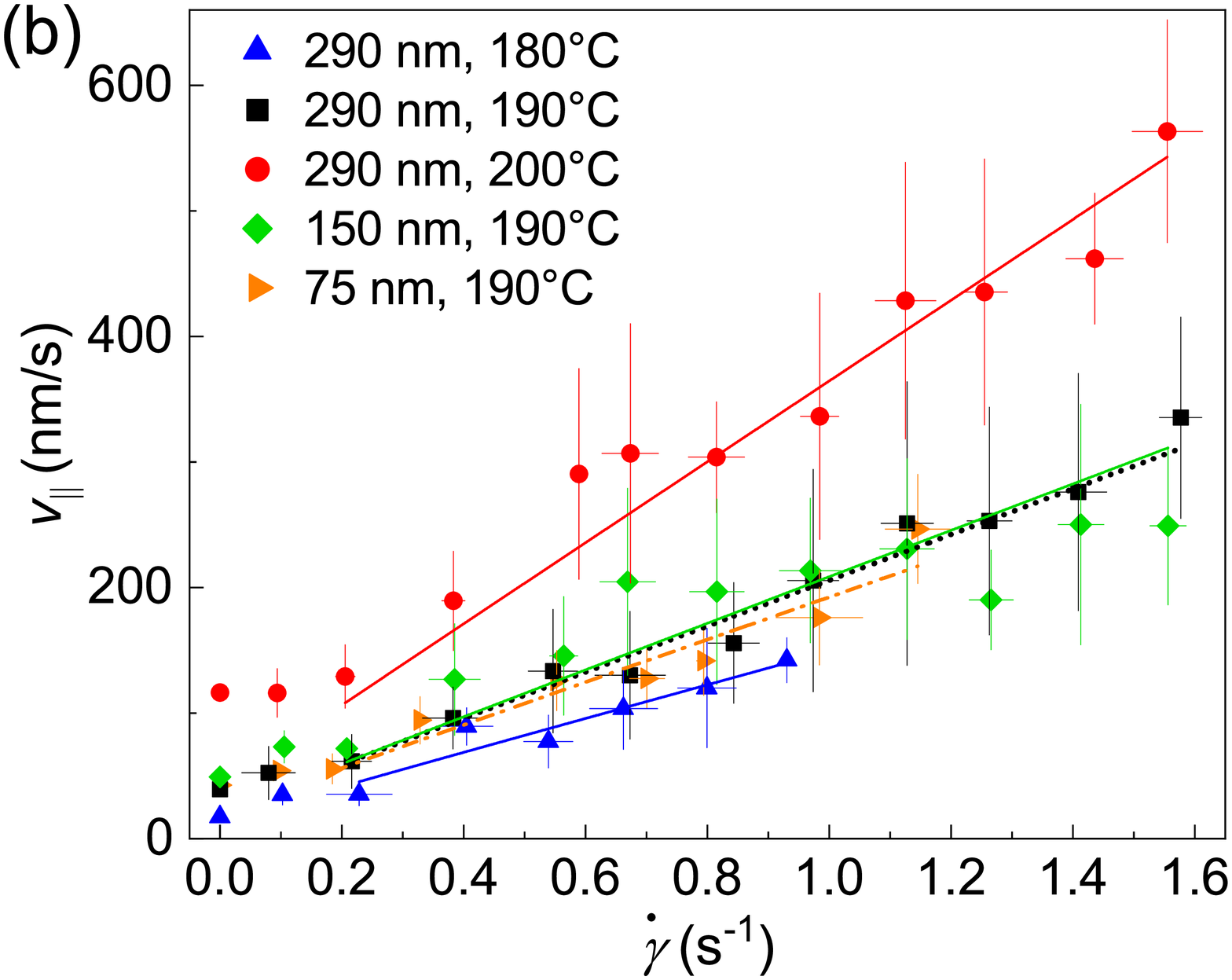}
  \caption{Growth speed perpendicular (a) and parallel (b) to shear as a function of shear rate. The lines represent the best fits using equations \ref{eq4} and \ref{eq8} respectively. The fits for $v_\parallel$ were obtained by fixing $A$ with its averaged value obtained from the fits for $v_\bot$.}
  \label{fig:model}
\end{figure*}

The obtained results are plotted as a function of shear rate in Figure \ref{fig:model}. Undoubtedly, for a given shear rate, $v_\bot$ is smaller than $v_\parallel$ and its increase with increasing shear is slower. Typically $v_\bot$ increased by a factor 2 from 0 to 1.6 s$^{-1}$, while the increase for $v_\parallel$ in the same interval is 5 to 8 times.

In the direction perpendicular to the applied shear, we first make the hypothesis that the growth speed can be described by the no-shear model (see Eq. \ref{eq3}), i.e. dewetting dynamics is only affected by shear parallel to the movement direction. 
However, we still must consider that PMMA displays shear-thinning behavior in the molten state, like most thermoplastics.

Hence the applied shear may modify its viscosity and consequently the dissipation in the vicinity of the rim.\\
We can then use the Carreau-Yasuda equation (see Experimental Section) to modify Eq. \ref{eq3} into:

\begin{equation}
v_\bot=\frac{Av_0}{\left[1+\left(\lambda\dot{\gamma}\right)^a\right]^\frac{n-1}{a}}
\label{eq4}
\end{equation}

Using the parameters obtained from the rheological measurements for PMMA (see Table \ref{tab:table1}), the best fits with only $A$ as a fitting parameter are presented in Figure \ref{fig:model}a. Though the data are somewhat scattered, the model captures reasonably well the trends for all conditions. Especially, at all temperatures the slight increase in growth speed with increasing shear rate can be described by the PMMA shear-thinning. Another experimental observation confirmed by the analytical model is that $v_\bot$ does not seem to depend on the initial PS film thickness (see green, orange and black points). Finally, all the values obtained for $A$ are close to each other, $A = 1.2 \pm 0.3$, in excellent agreement with $A_0$ obtained without shear. 

Let us now move to the growth speed along the applied shear direction $v_\parallel$. In this case there is a need to incorporate an additional term taking into account the external shear applied when balancing dissipated viscous power and the capillary energy changes:
\begin{equation}
\eta\rho^2R\left(\frac{v_\parallel}{\rho}\right)^2 \sim \sigma Rv_\parallel+\eta{\dot{\gamma}}^2\rho^2R
\label{eq5}
\end{equation}
where $\eta$ is, same as for $v_\bot$, a function of $\dot{\gamma}$. 

Reintroducing $A$ and another geometric constant $B$ for the term describing external shear, this leads to a solution of the form: 
\begin{equation}
v_\parallel=\frac{A\sigma}{2\eta}+\sqrt{{\frac{A^{2}\sigma^{2}}{4\eta^{2}}+B {\dot{\gamma}}^2}\rho^2}
\label{eq6}
\end{equation}

Here we can define a critical shear rate, $\dot{\gamma_c}$, such as:
\begin{equation}
\dot{\gamma_c} = \frac{\sigma}{\eta_0\rho} 
\label{eq7}
\end{equation}

Through factoring and expanding the square root we simplify the solution by identifying two regimes for $v_\parallel$:
\begin{equation}
  v_\parallel \approx
    \begin{cases}
    \scalebox{1.33}
      {$\frac{Av_0}{\left[1+\left(\lambda\dot{\gamma}\right)^a\right]^\frac{n-1}{a}}$}=v_\bot & \text{if $\dot{\gamma} \ll \dot{\gamma_\text{c}}$}\\
      \\
      \scalebox{1.33}
      {$\frac{Av_0}{2\left[1+\left(\lambda\dot{\gamma}\right)^a\right]^\frac{n-1}{a}}$}+ \sqrt{B}\dot{\gamma}\rho & \text{if $\dot{\gamma} \gg \dot{\gamma_\text{c}}$}\\
    \end{cases}
\label{eq8}
\end{equation}

To estimate the rim size, we use volume conservation and for the sake of simplicity will assume a toroidal shape. We also use an average value for the effective radius at a given set of parameters ($e,T$). With these hypotheses, we can write $\rho \approx \sqrt{e\langle\mathcal{R}\rangle/2\pi}$. \cite{Chebil2018}
It has been verified $\mathcal{R}$ indeed does not vary much from hole to hole nor during the limited experimental time nor with different shear rates (i.e. the morphology of the holes changes but their area does not). 
Coming back to Eq. \ref{eq7}, we obtain $\dot{\gamma}_\text{c}\lesssim 0.1$ s$^{-1}$ for all the systems studied. In Figure \ref{fig:model}b, we then use the equation from the second regime to fit the experimental data for $ 0.2 <\dot{\gamma} <1.6$ s$^{-1}$, with only $\sqrt{B}$ as a fitting parameter. The model predicts well the growth speed evolution with shear rate for every condition studied. The obtained values for $\sqrt{B}$ are slightly scattered, $\sqrt{B}=0.4\pm 0.2$, while a constant value, similarly to $A$, shall be expected. Still, the variations do not seem to follow a monotonic trend with either PS thickness or PMMA viscosity, and may then be due to the simplifying hypotheses to estimate $\rho$.\par 

\begin{figure}[t!]
  \centering
  \includegraphics[width=8.4cm]{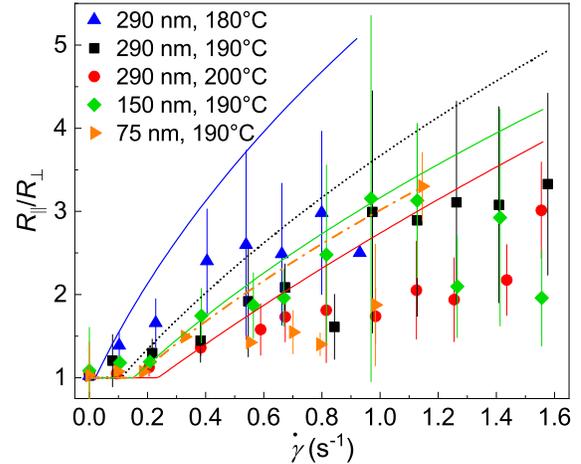}
  \caption{Radius ratio as a function of shear rate. The proportion of semi-major axis $R_\parallel$ divided by semi-minor axis $R_\bot$ was calculated similarly to the sets of data presented for the growth dynamics. 5 different holes were monitored during each experiment and minimum 5 images were analyzed for each hole. From the experimental setup, a shear rate value was obtained for each hole by measuring its moving speed. The data was grouped into intervals of 0.15 s$^{-1}$ and averaged over at least 5 points. The ratio over time remains constant during a particular experimental run and the average value for each hole is taken as a mean value from all analyzed images.}
  \label{fig:ratio}
\end{figure}

To further examine the validity of our model, we conclude by analyzing the ratio $R_{\parallel}/R_\bot$ as a function of the applied shear. The results are plotted in Figure \ref{fig:ratio}. Starting from a circular shape without shear (i.e. $R_{\parallel}/R_\bot=1$), the holes deform more and more as shear is applied (see also Figure \ref{fig:one}). Hence $R_{\parallel}/R_\bot$ increases for all temperatures and all PS thicknesses, up to a value close to 3 for shear rates of 1.6 s$^{-1}$.
Considering that this ratio should be equal to $v_\parallel$/$v_\bot$, the previous model can be used to describe it, following Eq. \ref{eq9}:
\begin{equation}
\frac{R_\parallel}{R_\bot} \approx
\begin{cases}
    1 & \\ \text{if $\dot{\gamma} \ll  \dot{\gamma_\text{c}}$}\\
      \\
      \scalebox{1.2}
      {$\frac{1}{2}\left[1+\frac{\sqrt{B}}{A}\frac{\dot{\gamma}}{\dot{\gamma_\text{c}}}\left[1+\left(\lambda\dot{\gamma}\right)^{a}\right]^{\frac{n-1}{a}}\right]$} & \\ \text{if $\dot{\gamma} \gg \dot{\gamma_\text{c}}$}\\
    \end{cases}
\label{eq9}
\end{equation}

Figure \ref{fig:ratio} shows the corresponding fits with the previously obtained values for $A$ and $B$, i.e. without new adjusting parameters. The model captures semi-quantitatively the data, especially small changes at low shear rates, as well as the sharp increase at higher ones. More specifically, experimental trends such as a higher ratio for higher PMMA viscosities or thicker PS films, are also well-described. \\
However, a slight overestimate of the predicted holes' elongation is observed in every cases. This could be due to the fact that shear plays a bigger role than only shear-thinning of the surrounding matrix in the dewetting dynamics perpendicular to it (see also Figure \ref{fig:model}a).

In conclusion, the quantification of dewetting dynamics of ultra-thin polymer films under shear was achieved for the first time, and reported for different temperatures and film thicknesses. Though shear can delay rupture \cite{Kadri2021, Davis2010}, once holes are created they grow faster with increasing shear. This was described satisfactorily by a simple analytical model balancing interfacial tension with the viscous dissipation occurring in the matrix, due to the rim's movement and the external shear applied. The model, assuming that shear does not affect the growth speed perpendicularly to the movement but only in the direction parallel to it, captures the experimentally obtained data. This study shall give insight to the instabilities happening during processes allowing the fabrication of such layered structures, such as multilayer coextrusion. Future work will aim at refining the proposed model. Especially, from a theoretical point of view we shall consider more precisely the effect of shear in the perpendicular direction, polymer-polymer interfacial slip, and focus experimentally on capturing more precisely the onset of dewetting as well as the shape of the rim. A set-up allowing to work at higher shear rates will also be of interest.

\section{Experimental Section}
Following our previous works \cite{Chebil2018,Zhu2016,Bironeau2017}, commercially available grades of PS and PMMA, namely PS 1340 (Total) and PMMA VM100 (Arkema), were selected for the study. The molecular weights and glass transition temperatures have been measured before \cite{Bironeau2017,Zhu2016}.

The viscoelastic properties of the polymers have been measured at three temperatures (180, 190 and 200\textdegree C) with a DHR 20 (TA Instruments) rheometer with a plate-plate geometry (25 mm diameter and 1 mm gap) under air flow. The experiments were performed on PS pellets and PMMA films samples with a similar thermal history to the trilayers. Frequency sweep tests were conducted in the range from 0.045 to 628 rad/s with an applied strain of 1\%, in the linear viscoelasticity region. The zero-shear viscosity ($\eta_0$), relaxation time ($\lambda$), and shear-thinning exponents ($a$ and $n$) were obtained for the two polymers at three temperatures using the classical Carreau-Yasuda model (see Eq. \ref{eq10}): \cite{Bird1968,Stadler2006}
\begin{equation}
|\eta^{*}\left(\omega\right)|=\eta_0\left[1+\left(\lambda\omega\right)^a\right]^\frac{n-1}{a}
\label{eq10}
\end{equation}
with $\eta^{*}$ the complex viscosity and $\omega$ the angular frequency in rad/s.\\
The obtained parameters are displayed in Table \ref{tab:table1} and show that PS and PMMA have a similar shear-thinning behavior. The analysis of the dewetting data was conducted assuming the classical Cox-Merz rule \cite{Cox1958}: $\eta(\dot{\gamma})=|\eta^{*}(\omega)|$.

\begin{table}[ht]
\caption{Parameters of Carreau-Yasuda fits}
\label{tab:table1}
\small
\begin{tabular}{cccccc}
\hline
\multirow{2}{*}{} & $T$ & $\eta_0$  & \centering{$\lambda$} & \multirow{2}{*}{\centering{$a$}} & \multirow{2}{*}{\centering{$n$}} \\

& (\textdegree C) & (Pa$\cdotp$s) & (s) & & \\
\hline
\multirow{3}{*}{PS} & 180 & 58400 & 2.96 & \multirow{3}{*}{0.91$^*$} & \multirow{3}{*}{0.37$^*$} \\
& 190 & 29200 & 1.48 & &\\
& 200 & 14600 & 0.74 & &\\
\hline
\multirow{3}{*}{PMMA} & 180 & 52800 & 0.91 & \multirow{3}{*}{0.75$^*$} & \multirow{3}{*}{0.34$^*$} \\
& 190 & 29200 & 0.43 & &\\
& 200 & 13100 & 0.15 & &\\
\hline
\multicolumn{6}{l}{$^*$ \scriptsize average values with standard deviations smaller than 0.01}
\end{tabular}
\end{table}

PS thin films were prepared via spin-coating (Spin 150 v-3, SPS) PS in toluene ($\geq$99.8\%, VWR) solutions onto a silicon wafer (100 crystal orientation, Sil’tronix) at 2000 rpm/s and 2000 rpm/s$^2$ for 60 s. Three previously filtered concentrations (1.5, 2.5, and 4.0 wt.\%) were used. Films with thicknesses $e=75\pm5$ nm, 150 $\pm$ 10 nm, and 290 $\pm$ 15 nm were obtained and controlled for each sample by AFM (Nanoscope V, Veeco) in tapping mode. \par

PMMA films were prepared by thermopressing pellets between two glass slides at 200\textdegree C in a laboratory press (Gibitre Instruments). Films with thicknesses from 70 to 170 $\mu$m were obtained, as determined by a digital length gauge system with accuracy of 0.2 $\mu$m (Heidenhain). The root mean square roughness values obtained by AFM was below 10 nm, in all cases considerably smaller than the PS thickness. Any possible impact of this roughness on the dewetting dynamics was neglected herein.\par
The PMMA films were sorted into pairs of similar thicknesses to ensure symmetric trilayers.
The PS films were cut into small pieces (about $10\times 10$ mm) and floated in a distilled water bath. A piece of PS film was then moved onto the PMMA substrate. The bilayer was left to dry at room temperature for about 12 h. Afterwards, the second PMMA film was placed on top. Such trilayer samples were kept in an oven at 60\textdegree C before the experiments to prevent water uptake.\par

Shearing experiments were conducted in a shearing hotstage (CSS450, Linkam Scientific Instruments) placed under an optical microscope (Eclipse LV150N, Nikon). The trilayer samples were placed on a quartz glass in the cell and pre-heated at 160\textdegree C for 3 minutes. Next, the gap, $H$, was set to 97.5\% of the trilayer's initial thickness and kept for another 3 minutes to ensure proper adhesion. Afterwards, the temperature was increased to the temperature of the measurement and the sample subjected to a chosen shear rate. Due to the limited observation window of the microscope and the movements induced by shear, this study only deals with $\dot{\gamma}<2$ s$^{-1}$. Such shear rates are nonetheless consistent with those typically achieved during multilayer coextrusion \cite{Bironeau2017,Beuguel2020}. \\ 

\begin{acknowledgement}
The authors thank Joshua McGraw and Frederic Restagno for fruitful discussions and critical reading of the manuscript. The Ecole Doctorale SMI (ED 432) is acknowledged for granting A.D. the fellowship for her PhD work. 
\end{acknowledgement}

\bibliography{biblio.bib}
\end{document}